\documentclass[superscriptaddress, reprint, amsmath, amssymb, aps, prx, floatfix]{revtex4-2}

\usepackage[utf8]{inputenc}
\usepackage{graphicx}
\usepackage{dcolumn}
\usepackage{bm}
\usepackage[normalem]{ulem} 
\usepackage{mathrsfs}
\usepackage{physics}
\usepackage{xcolor}
\usepackage{amsmath, amssymb, mathtools, amsthm}
\usepackage[colorlinks=true]{hyperref}

\usepackage{comment} 
\usepackage{orcidlink}

\usepackage{xr}

\graphicspath{{Figures/}} 

\definecolor{ve}{RGB}{0,161,22}
\definecolor{dmag}{rgb}{0.6,0.0,0.6}
\definecolor{pink}{rgb}{1,0,0.9}

\newcommand{\fixme}[1]{{\color{pink}#1}}    

\begin{abstract}
Spin transport typically relies on direct manipulation of the spin degree of freedom via magnetic fields, spin--orbit coupling, or engineered spin-dependent potentials. We show theoretically that directional spin currents can arise in a relatively simple setting---a one-dimensional interacting fermionic ring with static, spin-independent asymmetric barriers. By introducing asymmetric potential barrier geometry, spin-resolved circulating currents emerge on a closed chain even for symmetric initial configurations. The effect can be enhanced or reversed by appropriate initial state preparation and tuning the barrier asymmetry to resonant conditions.
\end{abstract}

\begin{document}



\title{Partially spin-polarized currents via tunneling through  asymmetric barriers}

\title{Towards spintronics via tunneling through asymmetric barriers}

\author{Elvira Bilokon\orcidlink{0009-0007-8296-2906}}
\email{ebilokon@tulane.edu}
\affiliation{Department of Physics and Engineering Physics, Tulane University, New Orleans, Louisiana 70118, United States}
\affiliation{Akhiezer Institute for Theoretical Physics, NSC KIPT, Akademichna 1, 61108 Kharkiv, Ukraine}

\author{Valeriia Bilokon\orcidlink{0009-0001-1891-0171}}
\email{vbilokon@tulane.edu}
\affiliation{Department of Physics and Engineering Physics, Tulane University, New Orleans, Louisiana 70118, United States}
\affiliation{Akhiezer Institute for Theoretical Physics, NSC KIPT, Akademichna 1, 61108 Kharkiv, Ukraine}

\author{Stanislava Litvinova}
\affiliation{Education and Research Institute ``School of Physics and Technology'', Karazin Kharkiv National University, Svobody Square 4, 61022 Kharkiv, Ukraine}

\author{Denys I. Bondar \orcidlink{0000-0002-3626-4804}}
\email{dbondar@tulane.edu}
\affiliation{Department of Physics and Engineering Physics, Tulane University, New Orleans, Louisiana 70118, United States}

\author{Andrii Sotnikov\orcidlink{0000-0002-3632-4790}}
\email{a\_sotnikov@kipt.kharkov.ua}
\affiliation{Akhiezer Institute for Theoretical Physics, NSC KIPT, Akademichna 1, 61108 Kharkiv, Ukraine}
\affiliation{Education and Research Institute ``School of Physics and Technology'', Karazin Kharkiv National University, Svobody Square 4, 61022 Kharkiv, Ukraine}

\date{\today}

\maketitle


\section{Introduction}
Controlling spin flow is a central goal of spintronics~\cite{Zutic2004, Awschalom2007, Smejkal2022, bazaliy2020interpretation, Fert2024}, with far-reaching implications for low-dissipation information processing and quantum technologies~\cite{Dieny2020, Yuan2022}. Spin transport is typically controlled by acting directly on spin—through magnetic fields~\cite{Johnson1985, Chetcuti2022}, spin--orbit coupling~\cite{Dyakonov1971, Sinova2015, Manchon2019}, or engineered spin-dependent potentials~\cite{Moodera1988, Arute2020}. This paradigm has shaped much of modern spintronics, where selectivity is encoded explicitly in the Hamiltonian. But can directed spin currents emerge from an external potential that treats both spin species identically?

In this work, we provide a positive answer to this question by demonstrating that a paradigmatic Fermi-Hubbard ring with spin imbalance and static, spin-independent asymmetric barriers is sufficient to generate spin-resolved net circulating transport. The mechanism requires only breaking of lattice inversion symmetry via asymmetric spin-independent barriers and precise initial-state engineering~\cite{Parsons2016, Chiu2018}. Moreover, building on our recent discovery of interaction-driven resonant tunneling through asymmetric barriers~\cite{Bilokon2025}, we show that tuning the barrier height ratio to the resonant condition unlocks a qualitatively new regime of counter-propagating spin currents, whose direction is fully controllable via initial state preparation.

Our findings identify a minimal Hamiltonian setting for interaction-assisted spin-selective transport~\cite{Brown2019} in programmable quantum matter, requiring no spin-dependent external ingredients. The proposed mechanism is directly realizable in state-of-the-art ring-shaped optical lattice experiments with cold atoms~\cite{Jordens2008, Bloch2008, Haller2015, Mazurenko2017, Cai2022}. More broadly, our results show that barrier asymmetry combined with controlled initial-state preparation suffices to transform a spin-blind potential into a spin-selective current generator. This establishes a new design paradigm in programmable quantum matter, where spin control emerges from geometry and state engineering alone.


\section{Model and Method}
\begin{figure}
    \centering
    \includegraphics[width=0.55\linewidth]{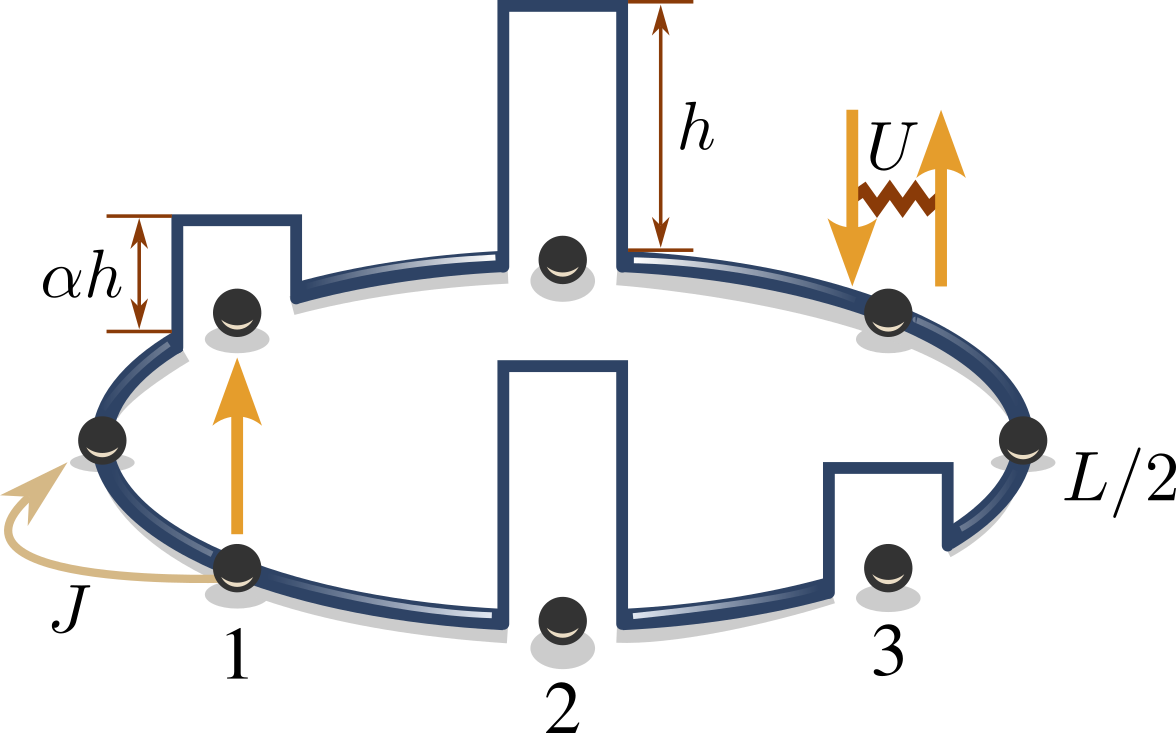}
    \caption{Schematic representation of the one-dimensional Fermi--Hubbard model with periodic boundary conditions and a site-dependent asymmetric external potential barrier of maximum height $h$. The hopping amplitude between neighboring sites is indicated by $J$, while $U$ represents the on-site interaction strength. Dark circles denote lattice sites.}
    \label{fig:model}
\end{figure}

We describe the system under study in the framework of the one-dimensional Fermi--Hubbard model with periodic boundary conditions (PBC),
\begin{eqnarray}
    \hat{\mathcal{H}} =
            -J \sum_{\langle i, j \rangle} \sum_{\sigma =  \uparrow,\downarrow}
            \left(
              \hat{c}_{i, \sigma}^{\dagger} \hat{c}^{}_{j, \sigma}
               + \text{H.c.}
           \right)
    \notag\\
    + \sum_{i=1}^L
           \left(
               U \hat{n}_{i, \uparrow} \hat{n}_{i, \downarrow}
                + h_i \hat{n}_{i}
           \right),
\end{eqnarray}
where $\langle i, j \rangle$ denotes summation over nearest-neighbor pairs, $L$ is the number of lattice sites, $J$ is the tunneling amplitude, and $U$ is the local interaction strength. The site-dependent (but state-independent) external potential $h_i$ describes the presence of asymmetric barriers ($\alpha$ is an asymmetry parameter) with specific heights at designated positions (see Fig.~\ref{fig:model}),
\begin{eqnarray}\label{eq:barrier}
    h_i = 
    \begin{cases}
    h,\quad & i=2, L/2+2;
    \\
    {\alpha h},\quad & i=3, L/2+3;\\
    0,\quad &\mbox{otherwise}.
    \end{cases}
\end{eqnarray}
The operator $\hat{c}^\dagger_{i,\sigma}$ ($\hat{c}_{i,\sigma}$) creates (annihilates) a fermion at site $i$ in the state with spin projection $\sigma \in \{\uparrow, \downarrow\}$, and $\hat{n}_{i,\sigma} = \hat{c}^\dagger_{i,\sigma} \hat{c}_{i,\sigma}$ is the corresponding number operator. The total particle number operator at site $i$ is $\hat{n}_i = \hat{n}_{i,\uparrow} + \hat{n}_{i,\downarrow}$.

The primary observable is the transferred charge. The total spin-resolved current operator is defined as
\begin{equation}
\hat{J}_\sigma = -iJ\sum_{\langle i, j \rangle} \left(\hat{c}_{i,\sigma}^\dagger \hat{c}_{j,\sigma} - \hat{c}_{j,\sigma}^\dagger \hat{c}_{i,\sigma}\right),
\end{equation}
where $J$ is the tunneling amplitude (here and below we use units $\hbar=1$). Hence, the transferred charge for spin $\sigma$ is computed as 
\begin{equation}
Q_\sigma(t) = \int_0^t \langle\hat{J}_\sigma(\tau)\rangle d\tau.
\end{equation}

The numerical simulations were performed using the QuSpin package~\cite{Weinberg2017, Weinberg2019}, a Python library designed for efficient exact diagonalization and time evolution of quantum many-body systems. 


\section{Results}

\subsection{Emergence of Directed Transport from Barrier Asymmetry}

We now focus on one-dimensional dynamics with periodic boundary conditions. To probe the emergence of spin-resolved transport in the presence of asymmetric potentials, we consider a ring of $L=8$ lattice sites populated by $N_\uparrow=2$ spin-up and $N_\downarrow=1$ spin-down fermions. The on-site interaction strength is fixed to $U=10J$, and the maximum barrier height is set to $h=20J$.

In this subsection, we restrict the system to a configuration in which the only source of asymmetry is the external barrier potential. Specifically, we prepare the initial state $\ket{\Psi(0)}$ as a superposition of Fock states, ensuring that the particles are distributed symmetrically between the barrier regions [see Eq.~\eqref{eq:barrier}] at $t=0$,
\begin{equation}\label{eq:symm_state}
    \ket{\Psi(0)} = \frac{1}{2} \Big(\ket{\uparrow}_1 + \ket{\uparrow}_8\Big) \otimes \Big(\ket{\uparrow\downarrow}_4 + \ket{\uparrow\downarrow}_5\Big).
\end{equation}

\begin{figure}[h!]
    \centering
    \includegraphics[width=\linewidth]{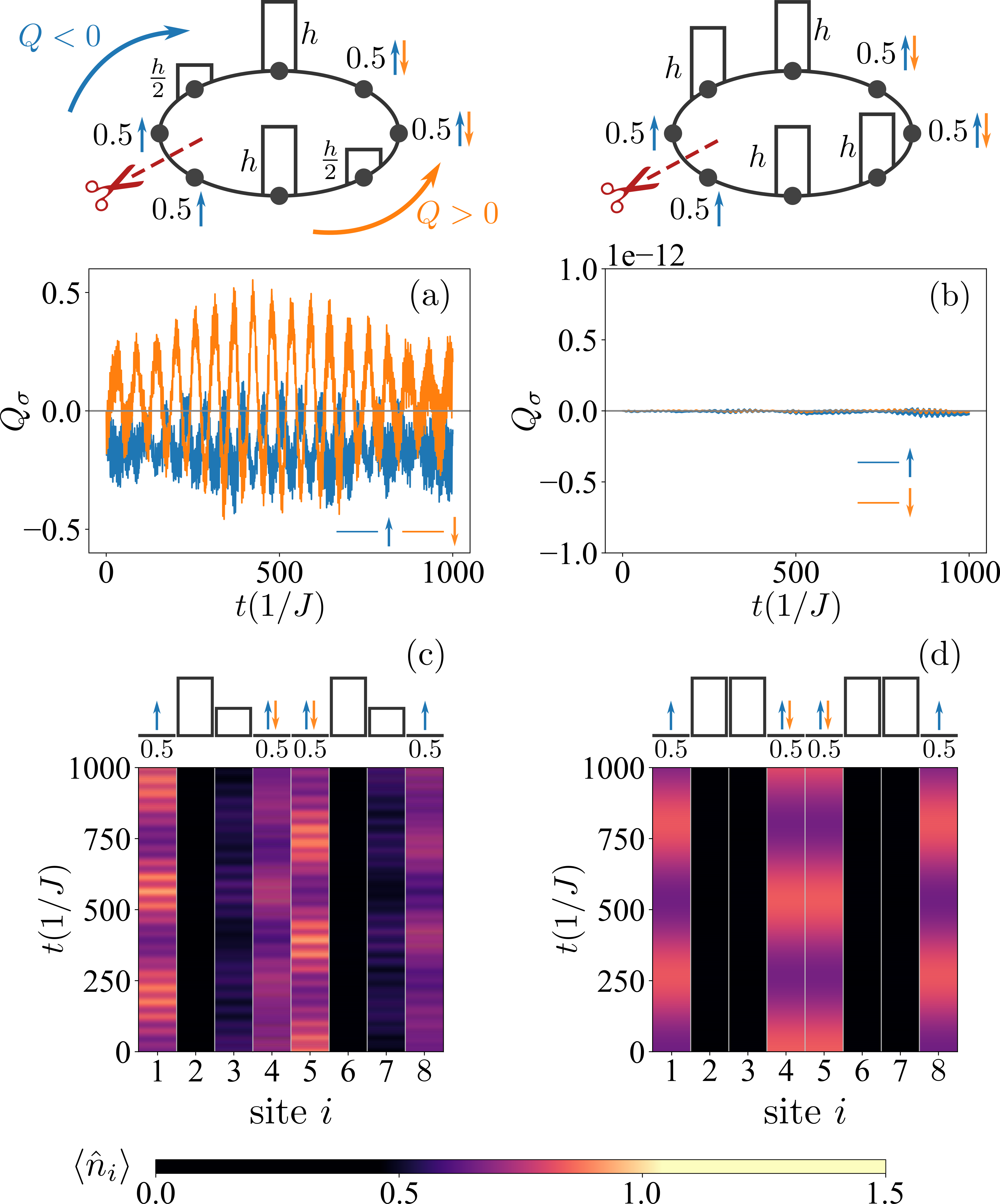}
    \caption{Barrier asymmetry as the essential requirement for directed spin transport. System dynamics for a symmetric initial state under asymmetric (left column, $\alpha = 0.5$) and symmetric (right column, $\alpha = 1$) barrier configurations. (a, b) Transferred charge for spin-up (blue) and spin-down (orange) components. A finite transferred charge develops only in the asymmetric case. (c, d) Site-resolved total density dynamics $\langle \hat n_i(t) \rangle$. The ring geometry is schematically unfolded above each panel, with the dashed cut in the top illustrations indicating where the periodic lattice is opened for visualization.}
    \label{fig:TransportFromAsymmetry}
\end{figure}
Figure~\ref{fig:TransportFromAsymmetry} compares transport dynamics for the initial state defined in Eq.~\eqref{eq:symm_state} under asymmetric and symmetric barrier configurations (both configurations are schematically illustrated on the top). In the left column with the asymmetric barriers, we observe a nonzero charge transfer for both spin components, demonstrating directed spin-dependent transport despite the absence of any initial spatial bias {for spin components}. The direction and magnitude of the transferred charge are uniquely determined by the barrier asymmetry. By contrast, in the right column when both barriers are set to the same height $h$, transferred charge vanishes up to numerical fluctuations (the same holds for lower values of $h$ and fixed $U$ and $J$). This comparison demonstrates that current generation relies critically on spatial asymmetry of the state-independent barrier potential, which provides the essential symmetry breaking required for directed transport.

To understand why symmetric barriers do not generate directed transport, we examine the site-resolved density dynamics $\langle \hat n_i(t) \rangle$ shown in Fig.~\ref{fig:TransportFromAsymmetry}. To guide the eye, the ring geometry is schematically unfolded above the density plots. The dashed cut on the rings (top panel of Fig.~\ref{fig:TransportFromAsymmetry}) indicates the point at which the periodic lattice is opened for visualization. In the asymmetric configuration, the density evolution breaks inversion symmetry and develops a preferred direction along the ring. Finite bond currents persist across all links, so that tunneling processes are not compensated by symmetry-related counterflows. As a consequence, local particle motion builds up into a net circulating current.

In contrast, the symmetric barrier configuration displays pronounced density oscillations while maintaining overall inversion symmetry. Although particles tunnel between neighboring sites, specific bonds exhibit strongly suppressed current, effectively partitioning the ring into dynamically weakly communicating segments.  As a result, motion persists locally but fails to establish a preferred circulation direction, and the net transferred charge remains zero.

Local density dynamics do not, by themselves, imply directed transport. A finite circulating current appears only when barrier asymmetry breaks lattice inversion symmetry and prevents cancellation of opposite tunneling processes.

\subsection{Modulating Directional Spin Transport by Initial State Preparation}
We demonstrated that a symmetric setup---a symmetric barrier combined with an initial state in which particles are positioned symmetrically with respect to the barriers---does not generate directed transport.
We now introduce a second source of symmetry breaking by preparing a spatially biased initial configuration, while keeping the barrier asymmetric and spin-independent.

\begin{figure}[h!]
    \centering
    \includegraphics[width=\linewidth]{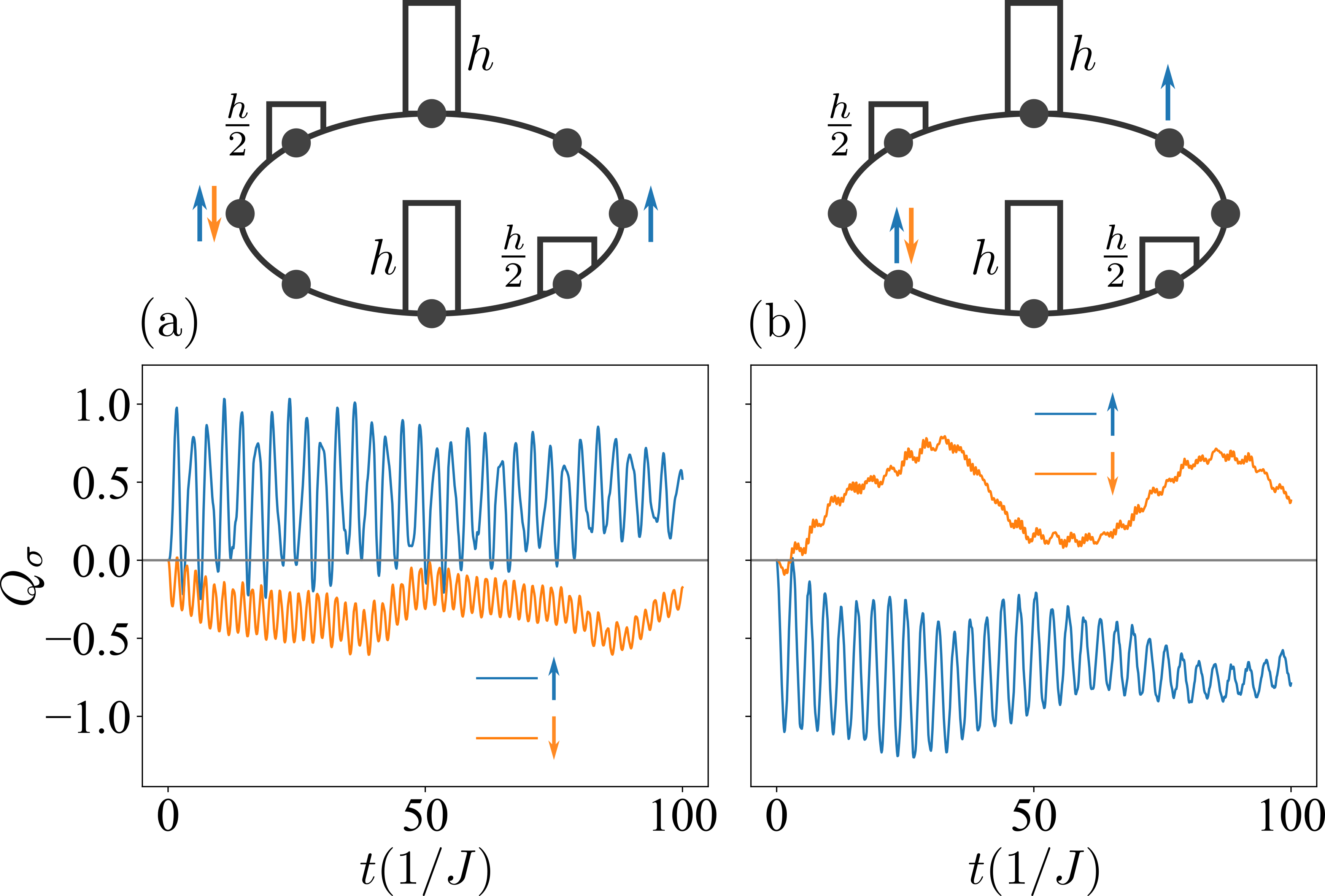}
    \caption{Transferred charge dynamics for spin-up~(blue) and spin-down~(orange) components under the asymmetric barrier configuration with two distinct initial states.  The doublon is positioned adjacent to (a) $h/2$ barrier and (b) $h$ barrier, with the unpaired spin-up fermion located three sites away along the ring.}
    \label{fig:SpinDirectionality}
\end{figure}

Figure~\ref{fig:SpinDirectionality} shows the dynamics of two distinct biased initial states under the same asymmetric barrier. In the first configuration (left panel), the doublon is positioned adjacent to the $h/2$ barrier, with the additional spin-up fermion on the opposite side of the ring. In the second configuration (right panel), the doublon is instead placed next to the higher $h$ barrier, while the remaining spin-up fermion is again on the opposite side. Experimentally, this can be realized using ultracold atoms in optical lattices~\cite{Greiner2002, Lewenstein2007, Jordens2008, Schneider2008, Bloch2008} by first loading and cooling atoms in an initial trapping potential, then switching on the asymmetric barrier potential to initiate the dynamics of interest~\cite{Greiner2002, Gericke2007}. 

The transferred charge clearly exhibits the opposite transport behavior in these two configurations. When the doublon is adjacent to the $h/2$ barrier [see Fig.~\ref{fig:SpinDirectionality}(a)], the spin-up component accumulates a positive transferred charge, whereas the spin-down component exhibits a transferred charge of opposite sign. Shifting the particles closer to the steeper side of the potential [see Fig.~\ref{fig:SpinDirectionality}(b)] reverses the pattern: the sign of the transferred charge flips for both spin species. In both cases, the transferred charge displays persistent oscillations throughout the evolution. 

This biased current behavior originates from the asymmetric mobility of the particles. The unpaired spin-up fermion constitutes the most mobile degree of freedom in the system and therefore initiates the dynamics by tunneling toward the neighboring empty site along the ring which is the energetically most favorable process. The spin-down fermion, initially bound within the doublon, moves in the opposite direction to form a doublon with the displaced spin-up particle. This interaction-driven rearrangement leads to a spin-down transfer opposite in sign to that of the spin-up component.

Importantly, directed transport can arise either from Hamiltonian asymmetry or from a spatially biased initial configuration. However, the counterpropagating motion of the two spin components strongly depends on {the presence of both} symmetry-breaking mechanisms. The interplay of these two asymmetries therefore provides a controllable route to spin separation: the barrier enables transport, while the initial state determines its direction.


\subsection{Control of Spin Current Directionality via Barrier Shape}

\begin{figure}[h!]
    \centering
    \includegraphics[width=0.95\linewidth]{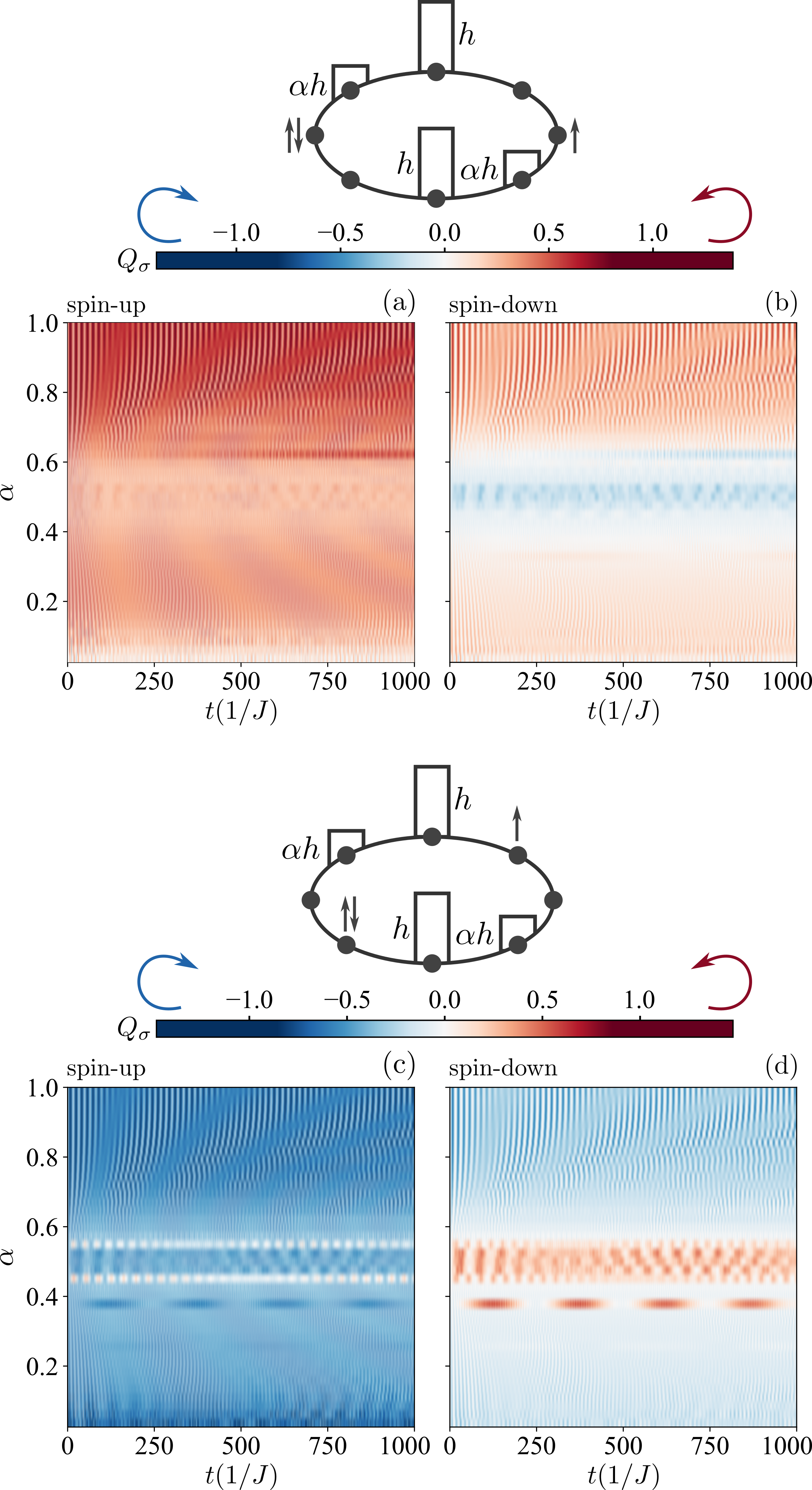}
    \caption{Control of spin directionality via barrier asymmetry. Transferred charge $Q_\sigma(t)$ as a function of time and asymmetry parameter $\alpha$ for two biased initial configurations (schematically shown above each panel set). Top row: doublon adjacent to the $\alpha h$ barrier. Bottom row: doublon adjacent to the fixed $h$ barrier. Left column corresponds to the spin-up component, right column to spin-down.}
    \label{fig:AsymmetryAlpha}
\end{figure}

To determine when two spin components propagate in opposite directions, we systematically vary the asymmetry parameter $\alpha$, which sets the height of one of the barriers [see Eq.~\eqref{eq:barrier}]. Figure~\ref{fig:AsymmetryAlpha} shows the transferred charge as a function of time and $\alpha$ for two biased initial configurations. 

For the configuration with the doublon adjacent to the $\alpha h$ barrier, the transferred charge of the spin-up component [Figure~\ref{fig:AsymmetryAlpha}(a)] remains positive over the entire range of $\alpha$. Conversely, when the doublon is positioned closer to the fixed $h$ barrier, spin-up fermions accumulate negative transferred charge throughout the evolution [Figure~\ref{fig:AsymmetryAlpha}(c)]. Thus, the direction of spin-up transport is dictated by the initial configuration and is largely insensitive to the degree of barrier asymmetry.

The behavior of the spin-down component, however, depends sensitively on the asymmetry parameter. For most values of $\alpha$, spin-down fermions move in the same direction as the spin-up particles---that is, both components are transported along the same circulation direction around the ring. A qualitatively different behavior appears only in a window around $\alpha\approx0.5$. In this regime, the spin-down transferred charge has opposite sign relative to the spin-up component, leading to counterpropagating spin motion. Remarkably, this occurs for both initial states. The condition $\alpha\approx0.5$ corresponds to $\alpha h\approx U$, i.e., when the lower barrier height becomes resonant with the interaction energy of the doublon.

These results establish that achieving two spin components being transported in opposite directions requires tuning the system into the resonant tunneling regime~\cite{Bilokon2025}. Outside this resonant regime, transport remains spin-polarized but co-propagating. We therefore conclude that counterpropagating spin transport depends not only on Hamiltonian asymmetry and initial-state bias, but also on resonant matching between barrier height~$h$ and interaction energy~$U$.

\section{Conclusion}
We have demonstrated that directional spin currents can emerge in a one-dimensional interacting Fermi--Hubbard ring driven by static, spin-independent asymmetric barriers. The first key ingredient enabling this behavior is barrier asymmetry. Spatially symmetric barrier configurations and initial state---prepared with particles distributed symmetrically with respect to the barrier positions---produce zero net transport. In contrast, introducing asymmetric barriers generates finite spin-resolved circulating currents even when the initial state itself contains no bias relative to the barrier geometry. A second, independent control parameter is provided by the preparation of the initial state, which determines the direction of spin flow. Crucially, counterpropagating spin currents are observed when these two symmetry-breaking mechanisms act together, and the barrier height is resonantly matched to the interaction energy, revealing resonant tunneling~\cite{Bilokon2025} as the key ingredient for spin separation. 

These results identify a rather straightforward, experimentally accessible route to spin control based on barrier geometry, interactions, and initial-state preparation. The mechanism can be implemented in ring-shaped optical lattices with programmable potentials~\cite{DelPace2022} and site-resolved state preparation, where both barrier asymmetry and particle configurations are routinely engineered. Extending this framework to time-modulated barriers may enable interaction-assisted atomtronic spin splitters and persistent circulating spin currents~\cite{Amico2022, Polo2025}, opening new directions toward geometry-driven spintronics in programmable quantum matter that operates without explicitly spin-dependent ingredients.

\textit{Acknowledgments.} 
This work was supported by Army Research Office (ARO) (grant W911NF-23-1-0288; program manager Dr.~James Joseph).
E.B. was supported by the National Science Foundation (NSF) IMPRESS-U Grant No.~2403609. A.S. acknowledges support by the National Research Foundation of Ukraine, project No.~2023.03/0073. The views and conclusions contained in this document are those of the authors and should not be interpreted as representing the official policies, either expressed or implied, of ARO, NSF, or the U.S. Government. The U.S. Government is authorized to reproduce and distribute reprints for Government purposes notwithstanding any copyright notation herein.

\bibliography{main}

\end{document}